\RequirePackage{lineno}

\documentclass[reprint, amsmath,amssymb,aps,
]{revtex4-2}

\usepackage{graphicx}
\usepackage{dcolumn}
\usepackage{bm}
\usepackage{hyperref}
\usepackage{braket}
\usepackage{physics}

\usepackage{color,soul}

\usepackage[dvipsnames]{xcolor}
\usepackage[normalem]{ulem}

\begin{document}

\preprint{APS/123-QED}

\title{Experimental wavelength-multiplexed entanglement-based quantum cryptography}

\thanks{Correspondence and requests for materials should be addressed
to Johannes Pseiner and Rupert Ursin.}%

\author{Johannes Pseiner}
 \email{johannes.pseiner@oeaw.ac.at}
\author{Lukas Achatz}
\author{Lukas Bulla}
\author{Martin Bohmann}
\author{Rupert Ursin}
 \email{rupert.ursin@oeaw.ac.at}
\affiliation{
 Institute for Quantum Optics and Quantum Information - Vienna (IQOQI), Austrian Academy of Sciences, Vienna, Austria\\
Vienna Center for Quantum Science and Technology (VCQ), Vienna, Austria}

\date{\today}

\begin{abstract}

   In state-of-the-art quantum key distribution (QKD) systems, the main limiting factor in increasing the key generation rate is the timing resolution in detecting photons.
   Here, we present and experimentally demonstrate a strategy to overcome this limitation, also for high-loss and long-distance implementations.
   We exploit the intrinsic wavelength correlations of entangled photons using wavelength multiplexing to generate a quantum secure key from polarization entanglement.
   The presented approach can be integrated into both  fiber- and satellite-based quantum-communication schemes, without any changes to most types of entanglement sources.
   This technique features a huge scaling potential allowing to increase the secure key rate by several orders of magnitude as compared to non-multiplexed schemes.
   
\end{abstract}

\maketitle

\section{\label{sec:level1}INTRODUCTION}

    In the last decades, the concepts of quantum key distribution (QKD) were introduced and its theoretical and experimental foundations have been developed \cite{wiesner,bennett,bennett2, ekert}. 
    These developments are driven by the goal of establishing secure point-to-point communication infrastructures \cite{hwang,ali, Hillery, gc} and eventually a quantum internet \cite{kimble, quantum_internet}.
    Ongoing scientific findings and technological innovations push the limits of QKD to ever-longer distances via optical fiber \cite{longdistance2,longdistancefiber, zbinden400, pan400km}, free-space \cite{scheidl}, and even satellites \cite{satellitetoground},
    connecting cities \cite{intercity} and continents \cite{satellite}, resulting in the first implementations of secure quantum networks \cite{Wengerowsky2018, florence,fieldtest,network, Chang2016, Joshi2019}.
    The main goal of to-date QKD research lies in increasing the key rates and developing loss- and noise-resistant \cite{eckernoise} strategies for long-distant quantum communication schemes which enables practicable and scalable long-distant QKD.

    The main limitation of modern QKD systems is not the generation at the source of the quantum states, but, in fact, lies in the detection of the photons.
    In all direct detection schemes, an assignment of two-photon detection at the respective receiver stations is done using a so-called coincidence timing window. 
    The minimal duration of this time interval is determined by the timing jitter of single-photon detectors and the readout electronics.
    Increasing the detection rate to the order of this timing limit---and hence increasing the probability of more than one photon per side within the window---unavoidably leads to an increase of the quantum bit error rate (QBER) eventually preventing key extraction at all \cite{steinlechner}. 

    A solution to this problem lies in entanglement-based QKD \cite{ekert, pan_realistic} by exploiting the inherent hypercorrelations naturally produced by most sources themselves. 
    Utilizing these correlations allows to deterministically separate wavelength correlated photon pairs into different detection channels.
    In this way, the timing limitation in the photon detection can be circumvented, leading to an intrinsically increased key rate.
    This is because the generation rate can be increased as the overall QBER is decreased and the source can be adjusted to operate at an optimized regime for the additional channels. 
    Other than deterministic separation, this beneficial behavior cannot be achieved by just probabilistically splitting the signals onto different detectors as in this case correlated detection events cannot be assigned accurately to each other.
    In fact, spectral multiplexing has revolutionized classical optical communication \cite{oldwdm} resulting in a leap in the classical communication capacity building the basis of modern telecommunication networks.
    The possibility of wavelength multiplexing has been adapted to the quantum realm where multiplexing of entangled states has been studied \cite{lim, aktas}.
    Using entanglement-based QKD, it is possible to exploit the tremendous potential of quantum state multiplexing for quantum communications.
    
\begin{figure}[t]
        \includegraphics[width=.95\columnwidth]{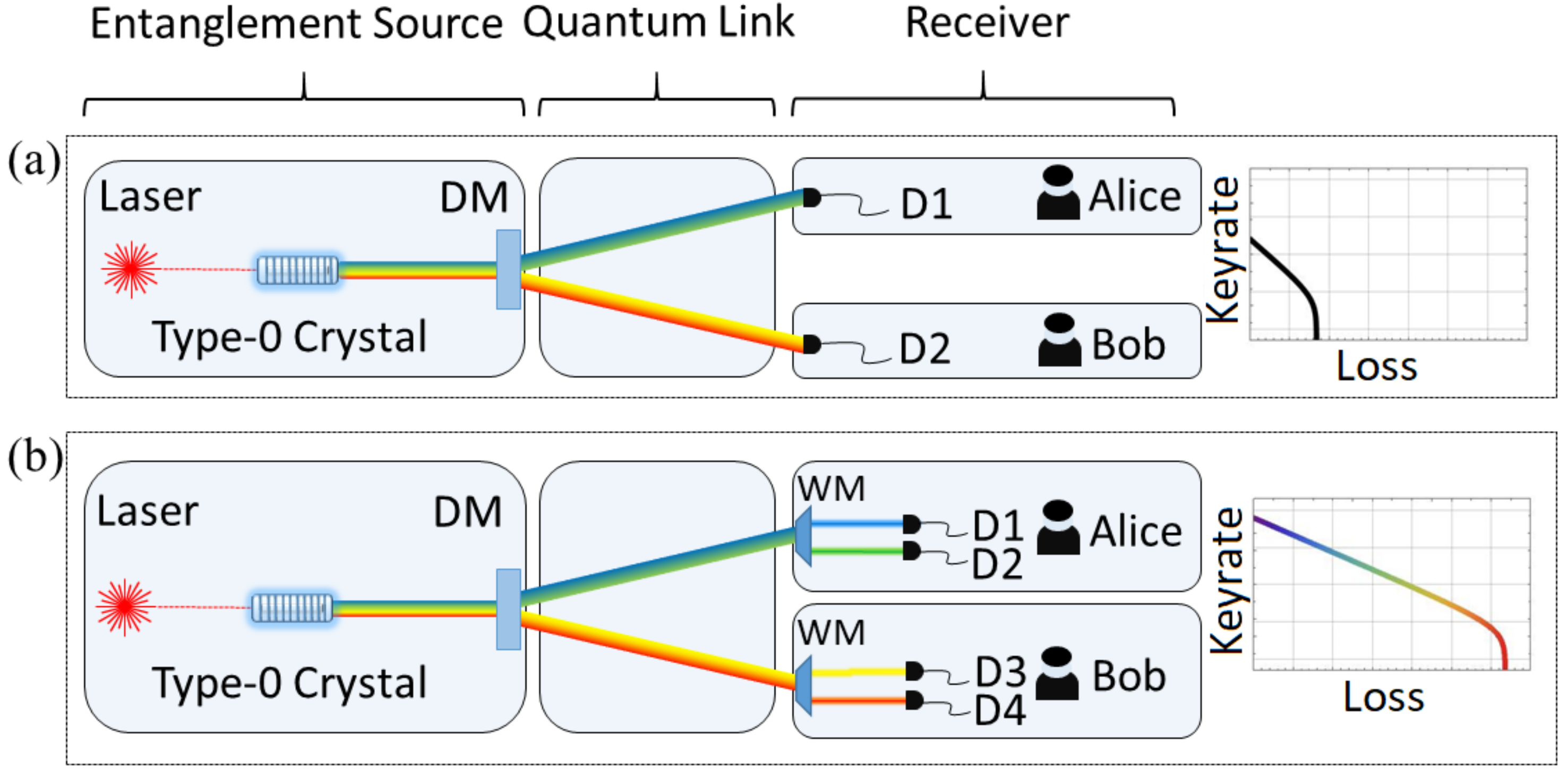}
        \caption{\label{fig:sketch} 
        Illustration of the implemented multiplexed QKD communication scheme.
        \textbf{(a)} Conventional entanglement-based QKD:
        entangled photon pairs are produced in a non-linear type-0 crystal, separated by e.g. a dichroic mirror (DM), and distributed via a quantum link to the receivers, Alice and Bob.
        A quantum secure key can be generated by measuring the polarization correlations of the photon pairs using the detection modules D1 (Alice) and D2 (Bob). 
        \textbf{(b)} Multiplexed entanglement-based QKD: 
        using wavelength multiplexing (WM), the number of channels can be increased leading to a significant speed-up in secure key rate.
        This idea is illustrated for two correlated channels [D1, D2 (Alice) and D3, D4 (Bob)], which can readily be extended to more channels.}
    \end{figure}

    Here, we present and implement for the first time a solution to overcome the major limitation of QKD systems posed by the timing inaccuracy in single-photon detection.
    Our scalable approach exploits the intrinsic wavelength correlations of entangled photon pairs which allows us to deterministically direct correlated photon pairs to corresponding polarization analyzation stages.
    In this way, the photon-pair rate can be increased drastically without saturation in the detectors, which results in an increased secure key rate and an effective decrease of the QBER.
    In our experiment, each end-user, Alice and Bob, receives one photon out of an entangled photon pair, analyzes and detects the photon pairs in the correlated wavelength channels and generates a secure key from the polarization measurement outcomes by treating each correlated channel separately.
    The source consists of an UV Laser, optically pumping a non-linear periodically poled KTiOPO4 crystal (see Fig. \ref{fig:sketch}), see Methods for details.
    The heart of our setup consists of a wavelength multiplexing (WM) system implemented at both receivers using volume holographic gratings (VHG) which reflect a spectral width of about $0.1\,$nm (46GHz frequency bandwidth) of the $4.7\,$nm (2.15THz) single photon spectrum.  
    By these means, our approach can fully harness the potential of bright single-photon pair sources, overcoming the limitations posed by the timing precision in photon detection.
    Furthermore, we demonstrate the scalability of this method which can be extended to the full photon-pair spectrum, and, hence, allows for increasing the attainable secret key rate by several orders of magnitude.
    Notably, the presented approach, which does not require any changes at the source, is compatible with fiber- and satellite-based communication and, thus, holds the potential to take global quantum communication to the next level.
    Adaptations of our scheme to other degrees of freedom are discussed.
    
\section{Method}

    \begin{figure*}
    \includegraphics[width=0.9\textwidth]{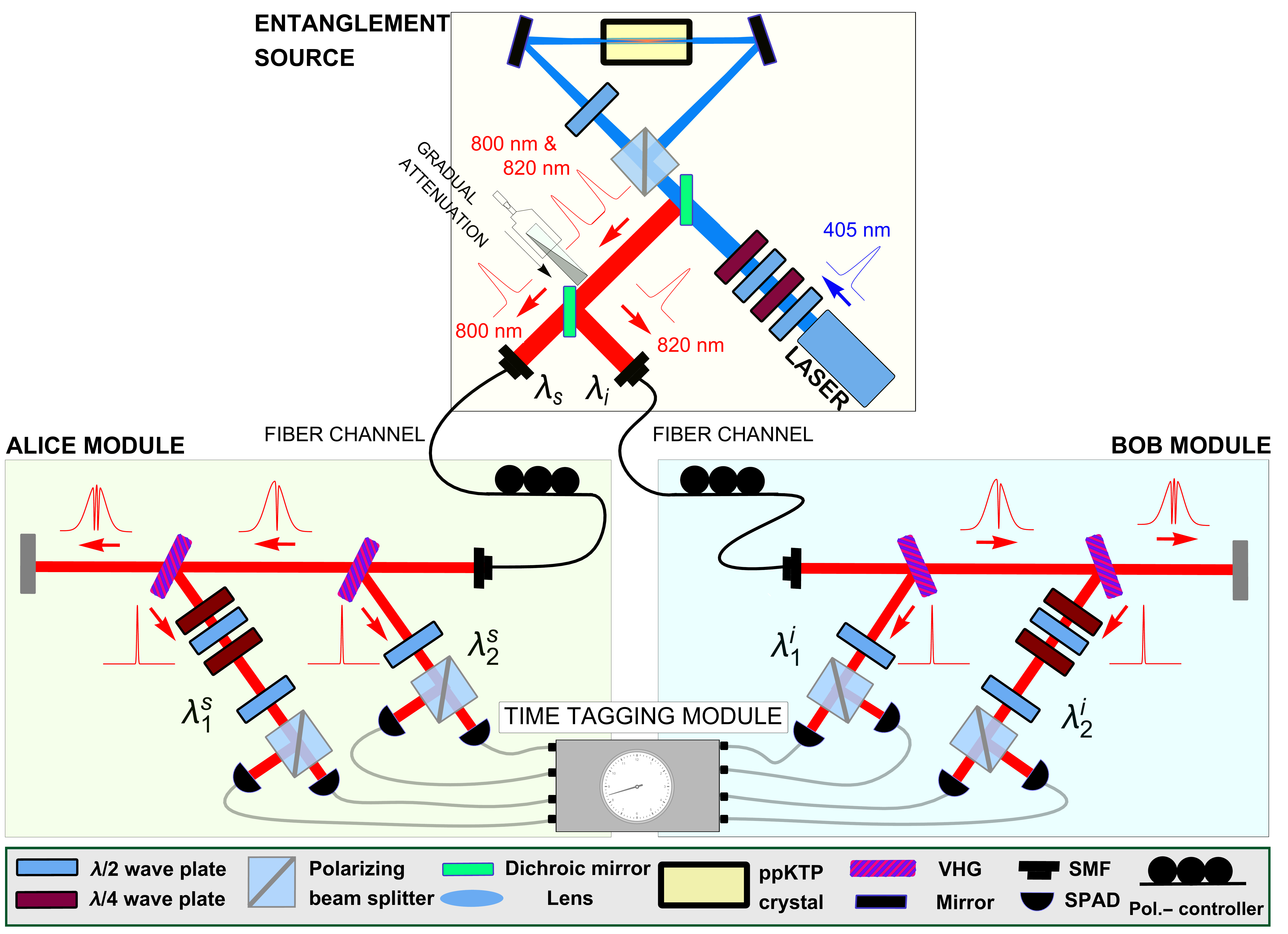}
    \caption{\label{fig:setup}
    Experimental setup of the entanglement source and the receiving sites Alice and Bob including the polarization-analysis stage and wavelength selection elements.
    The entangled SPDC photon pairs ($810\,$nm) were created in a Sagnac-loop configuration where a ppKTP nonlinear crystal was pumped bidirectionally by a continuous-wave pump laser ($405\,$nm).
    The photon pairs are sent to the communication partners via optical fiber channels.
    The VHG's located in the receiving sites perform a narrow-band wavelength-selection ($\sim \!0.12\,$nm/55GHz and $\sim \!0.24\,$nm/110GHz, for Alice and Bob respectively), followed by a polarization analysis.
    The count rates and time stamps were recorded by a time tagging module and correlations within a certain timing window $t_{c}$ (coincidence timing window) of around $1\,$ns were identified.
    }
    \end{figure*}
    
    \subsection{Details of the setup}
    
    A detailed outline of the setup is depicted in Fig. \ref{fig:setup}.
    The entangled photon pair source comprised a Sagnac-type configuration including a continuous-wave laser (grating stabilized $405\,$nm laser diode) producing photon pairs ($810\,$nm) via a spontaneous parametric down-conversion (SPDC) process through interaction with a type-0 periodically poled $3\,$cm-long KTiOPO4 nonlinear bulk crystal.
    The Sagnac-loop scheme provided photon pairs with high spatial and temporal indistinguishability and phase-stability, which in turn leads to a high degree of polarization entanglement.
    The quasi-phase-matching condition in a Type-0 nonlinear crystal featured a configuration in which the photon pairs propagate in the same temporal and spatial mode.
    The central wavelengths (CWL) of the SPDC photons were set by the temperature of the nonlinear crystal controlled by a Peltier element and a digital temperature controller.
    In this setup, a brightness of approximately $7.8\times 10^5\,$cps/mW of pump power after narrow-band filtering of $\sim \!0.1\,$nm (46GHz) was achieved.
    Finally, the results were recorded with an output pump power of $50\,$mW.
    
    Before separating the photon pairs, we introduced an attenuator simulating symmetrical channel loss (see Fig. \ref{fig:setup}). 
    Different attenuation levels were implemented through a gradient neutral-density filter on a motorized translation stage.
    The signal and idler modes were separated by a dichroic mirror with a cut-off wavelength close to the CWL of the SPDC spectrum ($810\,$nm).
    The CWL of the signal ($799\,$nm) and idler ($821\,$nm) photon spectra were chosen such that the wavelength-dependent splitting ratio of the dichroic mirror was sufficiently large, with a cut-off wavelength edge of around $20\,$nm (from $\sim$ \!5$\%$ to $\sim$ \!97$\%$ reflectivity with respect of the incident intensity).
    The separated signal and idler modes were coupled into single mode fibers acting as spatial filters.
    The polarization state of the photons was adjusted by in-fiber polarization controllers.
    The fibers were guided to the receivers, Alice and Bob, respectively.

    Using VHGs, signal and idler modes were multiplexed into two pairs of wavelength channels $\lambda^{\mathrm{s}}_1$ $\&$ $\lambda^{\mathrm{i}}_1$ (Ch. 1) and $\lambda^{\mathrm{s}}_2$ $\&$ $\lambda^{\mathrm{i}}_2$ (Ch. 2).
    The working principle of VHGs \cite{grating} is based on diffracting elements consisting of a periodic phase or absorption perturbation which allows for narrow wavelength-selective narrowband filtering based on the Bragg phase-matching condition.
    The VHGs used in the experiment reflect a spectral width with a FWHM of $0.12\,$nm (55GHz) (at Alice's site) and $0.24\,$nm (110GHz) (at Bob's site) of the total $4.7\,$nm (2.15THz) single photon spectrum [see Fig. \ref{fig:result} (a)].
    The respective CWLs of the channels and the mode spacing per site are depicted in Table \ref{tab:table1}.
    The spectra and the CWLs were recorded with a near-IR Ocean Optics QE65000 single photon spectrometer, with a spectral resolution of approximately $0.4\,$nm.
    Note, that the VHG's finite diffraction efficiencies ($\sim 70\%$ at Alice and $\sim 90\%$ at Bob) result in a respective transmission loss.
    
    \begin{table}
        \caption{
        \label{tab:table1}
        List of channel wavelengths and respective mode spacings.
        The mode spacings indicate the frequency spacings between the two color channels per user, hence between $\lambda_1^{\mathrm{s}}$ and $\lambda_2^{\mathrm{s}}$ and between $\lambda_1^{\mathrm{i}}$ and $\lambda_2^{\mathrm{i}}$.
        The spectral resolution of the spectrometer was approximately $0.4\,$nm.
        }
            \begin{ruledtabular}
                \begin{tabular}{ccc}
                \#Color Channel&Wavelength[nm]&Mode spacing[GHz]\\ \hline
                 $\lambda_1^{\mathrm{s}}$&$798.80\pm0.20$&$235\pm84.7$  \\
                  $\lambda_2^{\mathrm{s}}$&$799.32\pm0.20$&-\\
                  $\lambda_2^{\mathrm{i}}$&$821.31\pm0.20$&$239\pm89.0$\\
                  $\lambda_1^{\mathrm{i}}$&$820.77\pm0.20$&-\\
                \end{tabular}
            \end{ruledtabular}
    \end{table}
    
    The polarization-analysis modules for each channel comprised a half-wave plate for basis choice, followed by a polarizing beam splitter and silicon single-photon avalanche diodes for detection (Excelitas, Laser Components and PicoQuant APDs).
    For an independent polarization compensation in each channel, an additional retardation-plate system comprising of a quarter-, half- and quarter-wave plate was implemented.
    The detection efficiencies of all eight detectors were in the range of $50-70\%$ at $810\,$nm, while the dark count rate during the measurements never exceeded 1000 counts per second per detector.
    The timing jitter of the single-photon detectors plus electronics were in the range of 300-1000 ps.
    The recorded detection events were assigned with a time stamp provided by a time tagging module (TTM AIT 8000).
    Simultaneous clicks within a coincidence timing window $t_c$, which was chosen to be $1\,$ns, were identified as two-photon detections.
    
    \subsection{State characterisation}
    
    The  SPDC source produced polarization-entangled photon pairs in an anti-symmetric Bell state, $|\psi^-\rangle=\frac{1}{\sqrt{2}}(|H,V\rangle- |V,H\rangle)_\mathrm{pol}$.
    Due to energy conservation in the SPDC process, the frequency domain is entangled as well, leading to a hyperentangled state
    \begin{align}\label{eq:theory1}
        |\Psi^-\rangle=\int d\lambda\, c(\lambda)|\lambda_0+\lambda,\lambda_0-\lambda\rangle\otimes|\psi^-\rangle_{\mathrm{pol}},
    \end{align}
    where $\lambda_0$ is the central wavelength of the SPDC spectrum and $c(\lambda)$ is a continuous function of the wavelength $\lambda$ which characterizes the spectrum of the SPDC source and we assumed $\lambda\ll \lambda_0$.
    
    The spectral filtering allowed Alice and Bob to select different wavelength bands containing entangled photon pairs from the hyperentangled state in Eq. \eqref{eq:theory1}.
    Note that in this way, we utilized the wavelength correlations of the SPDC emission in order to implement wavelength demultiplexing in a deterministic way, while preserving the polarization entanglement between the considered photon pairs.
    In our experiment, we selected two wavelength-correlated photon pairs which corresponds to the following product state of two Bell states in polarization:
    
    \begin{align}\label{eq:theory2}
        |\psi^-\rangle_{\mathrm{MUX}}=\prod_{k=1}^2\otimes\frac{1}{\sqrt{2}}\left(|H_{\lambda_k^\mathrm{s}}V_{\lambda_k^\mathrm{i}}\rangle- |V_{\lambda_k^\mathrm{s}}H_{\lambda_k^\mathrm{i}}\rangle\right),
    \end{align} 
    
    where the labels indicate the wavelength channel of the signal $(\lambda_k^\mathrm{s})$ and idler photons $(\lambda_k^\mathrm{i})$ in the $k=1,2$ channels.
    
    To characterize the polarization correlations of the measured quantum states, we considered the state visibility $\mathcal{V}$ in two mutually unbiased bases (HV and DA basis in our case).
    The visibility in HV-basis adjusted to our experiment ($|\psi^-\rangle$ state) is given by
    
    \begin{equation}\label{eq:vis}
        \begin{aligned}
        \mathcal{V_{\mathrm{HV}}}  &=\frac{\text{maximum counts - erroneous counts}}{\text{total counts}}\\
        &=\frac{CC_{\mathrm{HV}}+CC_{\mathrm{VH}}-CC_{\mathrm{HH}}-CC_{\mathrm{VV}}}{CC_{\mathrm{HH}}+CC_{\mathrm{VV}}+CC_{\mathrm{HV}}+CC_{\mathrm{VH}}},
        \end{aligned}
    \end{equation}
    
    where $CC_{ij}$ is the total number of the recorded coincident counts in the HV basis, with the polarization settings at Alice ($i$) and Bob ($j$), respectively.
    The visibility for the DA basis is analogously defined.
    The visibilities are affected by erroneous counts due to experimental imperfections of the system and the false identification of photon pairs from uncorrelated wavelength channels.

    \subsection{Key rate extraction and data merging}
    
    For our entanglement-based QKD setup, an estimate for the attainable secure key rate can be calculated from our measurement results via \cite{maqkd}:

    \begin{equation}\label{eq:seckey}
        \begin{aligned}
        R_{\mathrm{s}}&=CC_{\mathrm{HV}}\cdot\frac{1}{2}\big(1-(1+f)H_2(Q_{\mathrm{HV}})\big)\\
        &+CC_{\mathrm{DA}}\cdot\frac{1}{2}\big(1-(1+f)H_2(Q_{\mathrm{DA}})\big)
        \end{aligned}
    \end{equation}
    
    where $CC_{\mathrm{HV}}$ ($CC_{\mathrm{DA}}$) is the total number of the recorded coincident counts in the HV (DA) basis and
    \begin{align}
        Q_{\mathrm{HV}}   &=\frac{\text{erroneous counts}}{\text{total counts}}=\frac{1-\mathcal{V_{\mathrm{HV}}} }{2},
    \end{align}

    is the quantum bit error rate (QBER) in the HV basis expressed in intensities of the polarization-analysis setting correlations, as defined as well in Eq. \eqref{eq:vis}, while the QBER in the DA basis $Q_{\mathrm{DA}}$ was calculated analogously.

    Furthermore,
    \begin{align}
    H_2(x)=-x \log_2(x)-(1-x)\log_2(1-x)
    \end{align}
    
    is the binary Shannon entropy and $f=1.1$ \cite{errorcorr} is the bi-directional error correction efficiency \cite{maqkd}.
    The basis choice was executed by setting the rotation angles of the HWP, while only one basis at the same time is recorded.
    
    We used a realistic key-rate model derived by Ma et al. \cite{maqkd} to analyze and compare the performance of our system.
    It included experimental parameters such as the link loss, the dark count probability and the brightness of the photon pair source.

    In order to compare the spectral resolving (WM) and the non-resolving (no WM) scenarios in terms of the key rate, we merged the detection events of the two corresponding detectors on Alice and Bob's side to mimic a single detector.
    This has been done by post-processing the detected events considering dead-time and the efficiency of the resulting detector. 
    The events were timestamped by the time-tagging module with a resolution of $1/12.15\,$GHz.
    By merging these recorded timestamps of two detectors as well as discarding events which are simultaneous and within a global dead-time, we received the detected events and physical properties of a single detector.
    
\section{RESULTS}
    
    \begin{figure*} 
        \begin{minipage}{\linewidth}
           \textbf{\hspace{0.3in}Experimental Data \hspace{2.5in} Scaling behavior}\par\medskip
            \centering
             \includegraphics[width=0.535\textwidth]{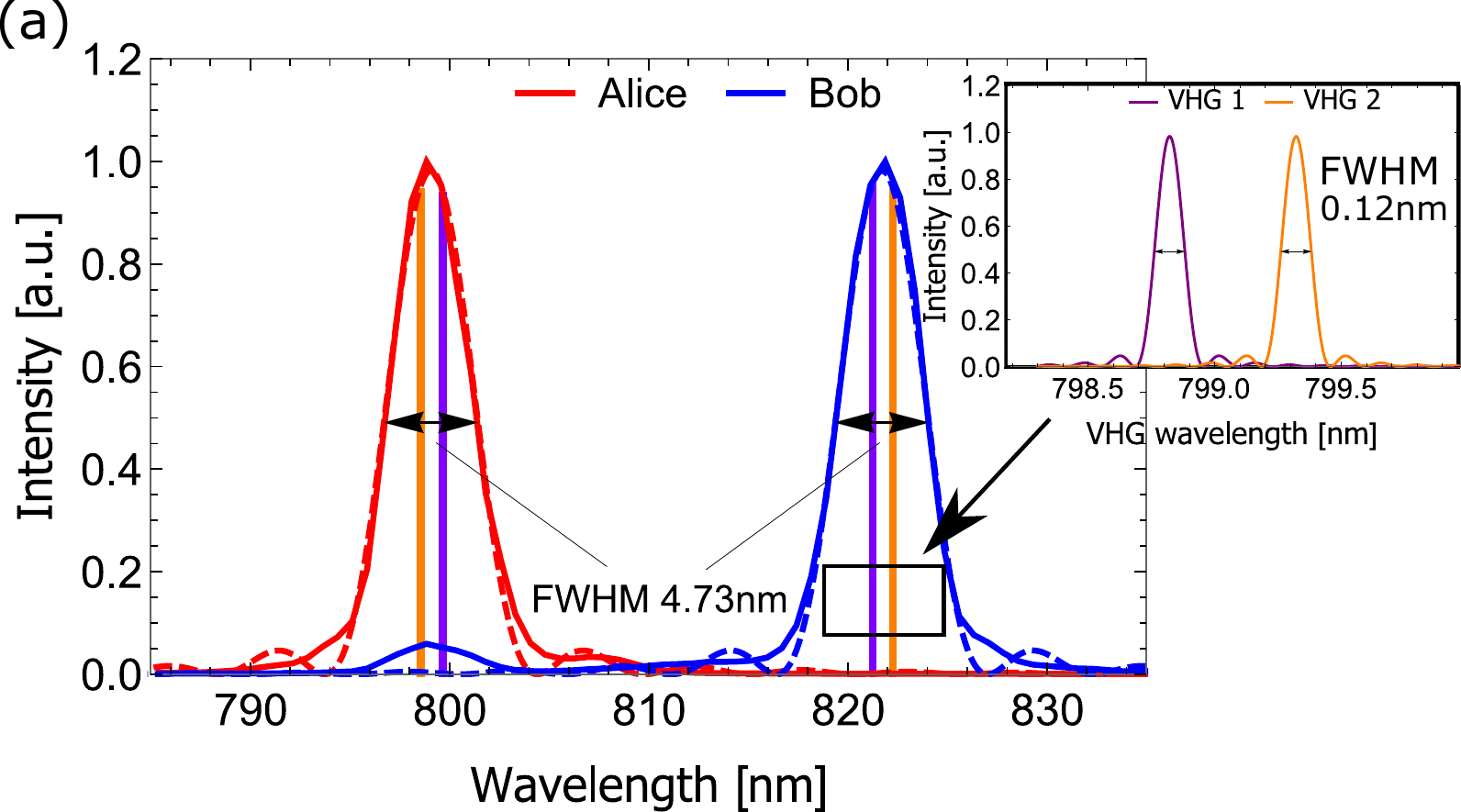}
             \includegraphics[width=0.42\textwidth]{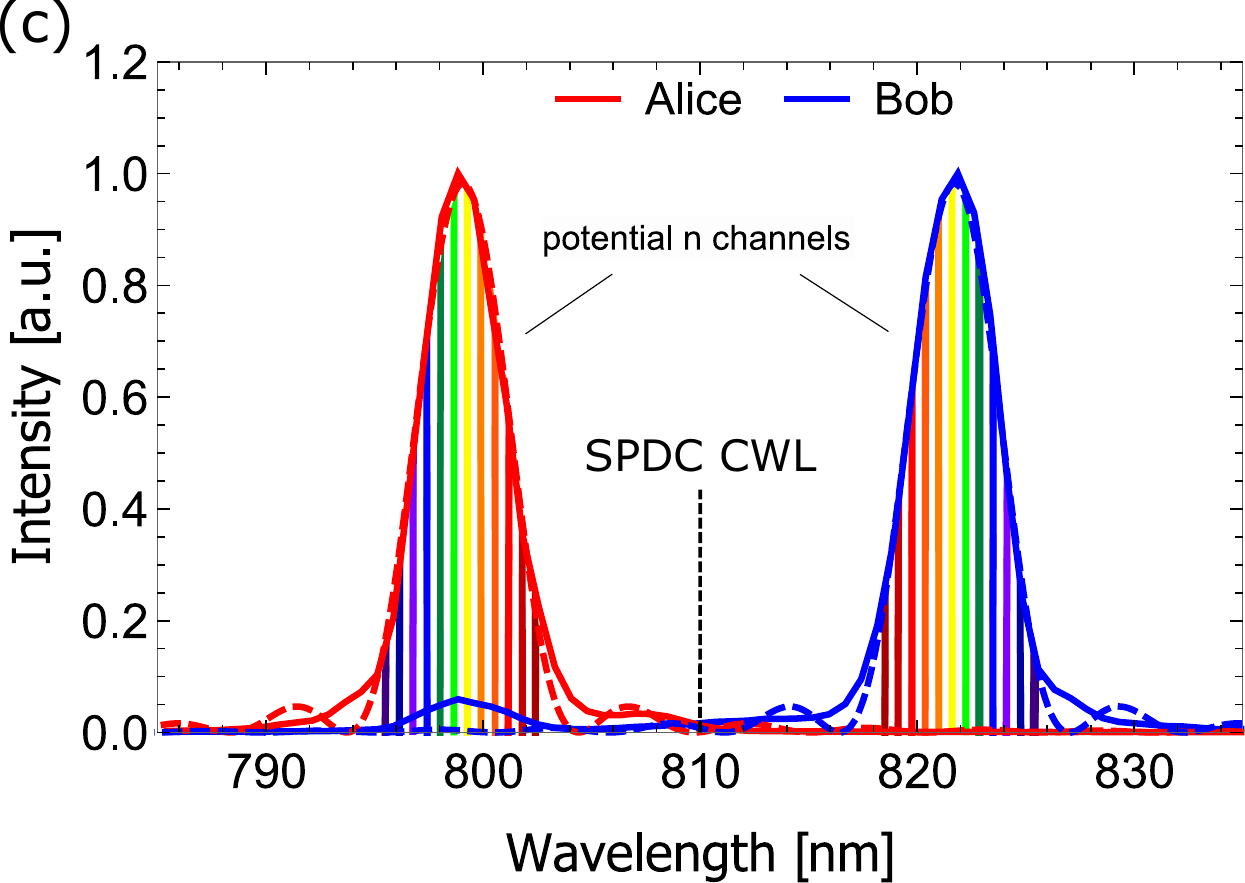}
        \end{minipage}
      \vspace{1em}
      \centering
             \hspace{-0.07in}
            \includegraphics[width=0.45\textwidth]{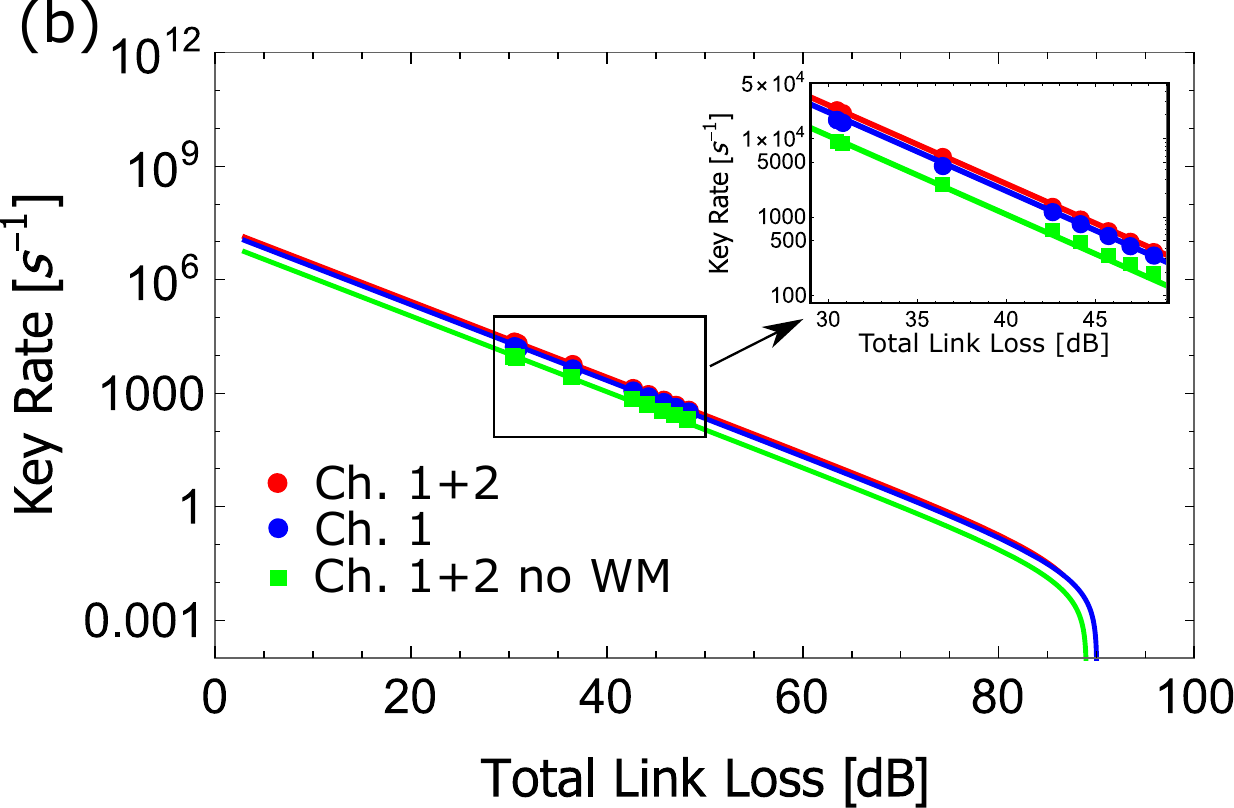}
             \hspace{0.53in}
            \includegraphics[width=0.45\textwidth]{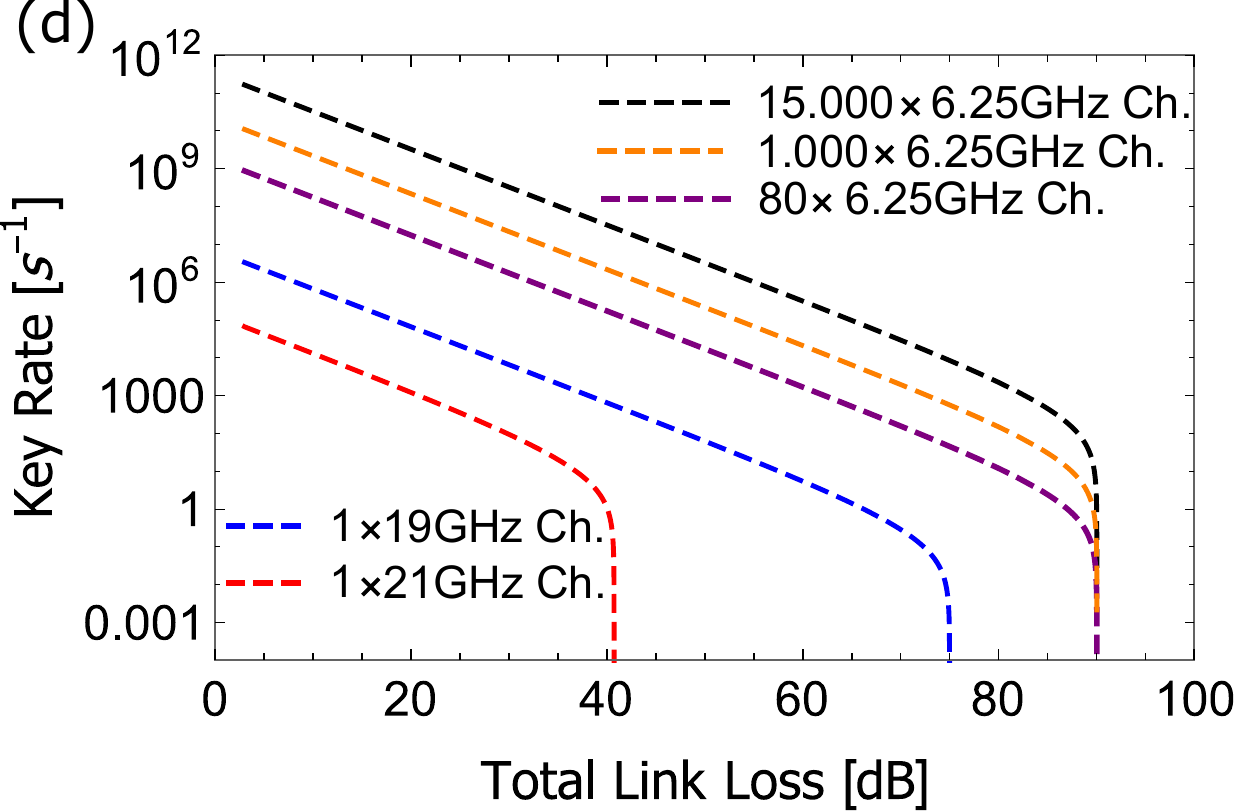}
        \caption{
        Representation of the experimental data of the implemented WM QKD scheme and its scaling behavior.
        \textbf{(a)} Normalized fit (dashed line) of the measured (straight line) intensity spectrum of the non-degenerate SPDC source is shown.
        Due to energy conservation, the signal (Alice) and idler (Bob) spectra have the same width with a full width at half maximum (FWHM) of $\sim \!4.73\,$nm.
        In addition, the parts of the spectrum selected by the VHGs are depicted, with a FWHM of $\sim \!0.12\,$nm (55GHz) and $\sim \!0.24\,$nm (110GHz) (provided by the manufacturer) at Alice and Bob's sides, respectively.
        \textbf{(b)} Secure key rates generated by our system is based on polarization measurements are shown for different loss levels.
        The dots represent the key rates generated from one (Ch. 1) and two (Ch. 2) combined WM channels exploiting the wavelength anti-correlation of SPDC photon pairs.
        The green squares correspond to the case in which both wavelength channels are jointly recorded (Ch. 1+2 no WM).
        This leads to the elimination of coincidence events which result in an increased QBER and a reduction in secure key rate. 
        \textbf{(c)} Illustration of the scalability potential of the presented approach by using more than two channels.
        The colors correspond to the frequency correlations with respect to the central wavelength (CWL) of the SPDC spectrum ($810$ nm).
        Both depictions of the spectra were normalized to the stronger signal for displaying simplicity.
        \textbf{(d)} The dashed lines indicate the scaling potential of our WM-featured QKD scheme using $80,1000$ and $15000$ wavelength channels of our source.
        A sufficiently broad spectrum of the photon pairs would allow for $n=15000$ wavelength channels, when taking optical transmission window of the atmosphere and a spectrum filtering of $6.25\,$GHz into consideration, see Sec. \ref{sec:Scalability} Scalability for more details.
        Additionally, the behavior for two broader wavelength channels with bandwidths of $19$ GHz and $21$ GHz is shown.
        }
        \label{fig:result}
    \end{figure*}
    
\subsection{Achieved key rates}

    The secure key rates are calculated based on the measured coincident counts [Eq. \eqref{eq:seckey}] of the correlated wavelength channels individually; see Method for details.
    In Fig. \ref{fig:result} (b), the measured secure key rates for different loss values are plotted together with the theoretical model, as introduced in the Method section.
    We start by analyzing the case in which Alice and Bob treat each wavelength channel separately. 
    From the obtained coincidence data, the communication partners can then extract two keys which can be added to obtain the overall key in this configuration (Ch. 1+2).
    We compare the obtained results with the case in which no WM is implemented (Ch. 1+2 no WM), i.e., that the two wavelength channels are not recorded separately but jointly measured using only one polarization analyzer at each receiver side [see Fig. \ref{fig:result} (b)].
    
    One would expect that in both cases the overall key rate is the same as the number of photon pairs transmitted and registered is equal.
    Our results show a considerable difference between the two cases.
    In fact, by using WM we obtain a higher key rate as compared to the joint-measurement case for all loss levels and we achieve an up to 2.6-times higher key rate for the lowest loss case.
    Even more surprising is that just one wavelength channel alone [Ch. 1 in Fig. \ref{fig:result} (b)] yields a higher secure key rate (by a factor of 1.9) than the joint measurement case although one channel only receives half the number of photons compared to the joint case.
    This is the reason why the overall key rate with WM is more than a factor of two larger than the corresponding key rate without multiplexing.
    It is crucial to mention, that the choice of the mean photon pair number of the individual wavelength channels close to the optimum value results in the largest improvement factor compared to non-multiplexed scenario (in our case, by a factor of 6.76).
    For high losses ($\sim \!89\,$dB), working at the saturation level of the detectors in time, one can still distil a secret key whereas Ch. 1+2 no WM does not yield any key.
    The reduced accidental rate due to the multiplexing manifests itself in the high loss regime ($\sim 89$dB), where the multiplexed system yields a non-vanishing key compared to the non-multiplexed system, effectively caused by a reduced QBER.
    Crucially, the steep decrease in the key rate at maximal tolerable loss and the overall key rate performance differs also within the two scenarios due to the superior systematic visibility (i.e., experimental imperfections) in Ch. 1 compared to Ch. 1+2 no WM. 
    Note, that the steep deterioration in key rate in the highest loss regime represents a well-known behavior for standard QKD systems \cite{neumann, scheidl, satellite, satellitetoground}.

    The difference between the two detection scenarios can be understood by considering the recorded coincidence-count rates and the resulting QBER.
    In the case where the two wavelength channels are not analyzed individually, the detection rate of each detector is higher.
    This increases the probability of detecting uncorrelated photons, i.e., photons from different wavelength channels, within the coincidence timing window.
    These events are so-called accidental coincidences which result in an increase in QBER and in turn negatively affect the secure key rate.
    In the multiplexed channels, we achieve a QBER for the lowest link loss ($30\,$dB) of 3.8$\%$ for one channel while the QBER of the joint measurement case is as high as 7.6$\%$.
    Note that a QBER lower than 9.5$\%$ is necessary to obtain a positive secure key rate \cite{gc}.

    The only possibility to avoid this detrimental effect would be to reduce the overall photon-pair production rate of the SPDC source. 
    In this way, the probability of falsely assigning uncorrelated photons could be reduced, however, at the cost of a lower pair and thus key rate.
    Hence, this result clearly shows that state-of-the-art QKD experiments are limited by the timing capability of the detection stage rather than by the source's production rate.
    And more importantly, our experiment provides an elegant way of overcoming this limitation by making use of the intrinsic wavelength correlations of the entangled photons and, hence, fully exploits another potential of entangled-photon sources.
    This advantage is based on the deterministic separation and individual detection of the correlated wavelength bands.
    In this way, the overall photon-pair rate can be increased while the timing saturation of the detectors is avoided and the number of accidentally assigning uncorrelated photon pairs to each other decreases.
    Consequently, implementing WM allows to raise the overall attainable secure key rate of entanglement-based QKD systems to a next level. 
    It is crucial to stress that this approach only requires a subtle modification at the receivers without changing the source.
    
\subsection{Scalability}
\label{sec:Scalability}
    
    By using $n$ wavelength channels one can improve the secure key rate linearly by up to several orders of magnitude, limited only by the bandwidth of the entangled photons and the available wavelength selection bandwidth. 
    Figure \ref{fig:result} (a) shows the fit of the measured SPDC spectrum of our source together with the four wavelength bands which we selected with the VHGs.
    Both signal and idler spectra have a FWHM of $4.73\,$nm and the selected spectral bands have a width of approximately $0.1\,$nm (46GHz).
    The filtering is based on reflection and the remaining signal bandwidth is transmitted rather than dumped, which presents a huge scaling capability of the presented approach as in this way the whole SPDC spectrum can be used (see Methods for details of VHG's working principle).

    In principle, it is possible to implement WM to use $n$ correlated channels which corresponds to the transmission the overall state $|\psi^-\rangle_{\mathrm{tot}}= \prod_{j=1}^n\otimes |\psi^-_{j}\rangle$ where each $|\psi^-_{j}\rangle$ represents a polarization entangled state.
    If we assume for simplicity that each channel has the same properties (intensity, detector characteristics, polarization compensation, dark count probability), this approach would give us a $n$-times higher key rate with respect to a comparable optimal single-channel setting (no WM).
    Note that without the multiplexed detection one would saturate the detectors in time and it would be impossible to generate a key in this scenario.
    This effect is illustrated in Fig. \ref{fig:result} (d) where the key rates for broader channels $19\,$GHz and $21\,$GHz---with respect to the optimal bandwidth $6.25\,$GHz as discussed in the scaling behavior---are shown.
    For these cases, we obtain a comparably low key rate and a reduced maximal amount of loss for which it is still possible to extract a key, while for $22\,$GHz bandwidth and higher, no key can be extracted any longer.
    
    A commercially available DWDM (dense wavelength division multiplexing) solution provides the possibility to utilize 80 wavelength channels for optical communication schemes \cite{80chan}, which can be implemented in QKD systems as well.
    Note that laser and crystal design allows for the optimized generation for all channels simultaneously \cite{raicol} and with state-of-the-art technology it is possible to realize multiplexing into $1000$ channels \cite{Ohara_1000_channels,takara_1000_channels}, already today.
    Moreover, a scalability argument for optical free-space long-distance setups can be given by considering the electromagnetic absorption within the atmosphere.
    Given future laser and crystal design, the absorption band between the oxygen A-band at $761\,$nm \cite{oxygenband} and the water vapor absorption band at $970\,$nm \cite{h2o} would allow for more than $n$=15000 wavelength channels assuming a $6.25\,$GHz bandwidth \cite{Ohara_1000_channels}.
    The separation of these wavelength bands can be achieved via a stacked arrangement of VHGs or implementing UDWM systems and subsequent polarization analysis of each band.
    The resulting boost in the overall key rate by using $n=80, 1000\textrm{ and }15000$ multiplexing channels is depicted in Fig. \ref{fig:result} (d).
    In the latter case, a geostationary dual downlink satellite mission with a total link loss of $70\,$dB could reach a key rate of $3\times10^4$ bits per second, assuming sufficiently large telescopes as well as low background noise.
    Note, that the theoretical photon pair rate was adjusted close to an optimal value of the QBER.
    
    The beneficial effect of WM is fundamentally limited by the increase of the photon's coherence time due to narrow-band spectral filtering.
    However, this is not an issue for implementations with state-of-the-art technology.
    For example, ultra-dense wavelength division multiplexing (UDWDM) allows for wavelength separation with a mode spacing of $6.25\,$GHz \cite{Ohara_1000_channels}, which results in a coherence time of $\sim \!50\,$ps, which is still much lower than typical coincidence timing windows, which are limited by the detector jitter and readout jitter to $\sim \!1\,$ns.
    
    Summarizing, by exploiting intrinsic properties of the entangled photon pairs which only requires an adapted architecture at the receiver sites, the secure key rate can be increased drastically.
    
\section{DISCUSSION}

    In this experiment, we successfully demonstrated an improvement of the quantum communication key rate through wavelength multiplexing for the first time. 
    We achieved an increase of the secure key rate by exploiting two correlated wavelength channels of our entangled photon source.
    We find that the speed-up increases linearly with the number of multiplexing channels employed, limited only by the damage threshold in the pump field of the nonlinear crystal (above which the crystal is critically damaged) as well as the bandwidth of the entangled photons and the available wavelength selection width.
    In that way, it allows for an improvement of several orders of magnitude compared to the case without WM.
    Specifically, in a regime where the photon-pair source would saturate the detectors in time, the WM scheme leads to a two-fold improvement: accidental coincidences are decreased, resulting in a reduced QBER per channel, and multiple-channel operation, enabling scalability with the number of channels, is achieved.
    Notably, the speed-up through channel multiplexing is not limited to the wavelength degree of freedom (DOF) but can be extended to any degree of freedom in the system implemented.
    In the case of photons, one could, for example, also use spatial correlations \cite{Walborn2010, ortega2021experimental} or orbital angular momentum \cite{krenn_oam} for this purpose.
    Moreover, the single photon's capacity of information can be exploited, too, thus presenting another promising method to increase the key rate in QKD protocols. To this end, one can utilize high-dimensional encoding in one DOF or use the hyper-encoding in multiple DOFs \cite{Yan2020, dellantonio, mower, hu}. 

    Integration of the proposed technique in existing and future quantum-communication links--- optical fiber and free space---is possible.
    Using fiber connections, one can harness well-established fiber-based WDM systems, which allows a direct integration into already existing classical telecommunication systems.
    Furthermore, our setup is capable to cope with the detrimental influences of atmospheric turbulence \cite{vasylyev}, hence allowing implementations over long-distance free-space links.
    In particular, the acceptance angle of the VHGs ($\sim \!900\!\!$ $\mu$rad) is larger than the angle-of-arrival fluctuations in typical (downlink) satellite-based long distance experiments due to atmospheric turbulences of about $10\!$ $\mu$rad \cite{aoa}.
    Considering the optical transmission window of the atmosphere, sufficiently broad quantum sources would allow for $n=15000$ wavelength channels.
    An additional advantage of the proposed technique is that the narrowly filtered detection bands provide a drastically increased ratio between signal and environmental background radiation.
    Thus, our WM scheme is intrinsically more robust against background radiation which is becoming of particular importance in daylight QKD applications.
    
    Note, even if the detectors performance technologically improves in the future drastically \cite{Najafi2015}, our system still features a high degree of scalability compared to non-multiplexed QKD systems.
    We note, however, that the limit in distance does not change when using more channels as each correlated wavelength channel pair can be seen as an independent sources and all detectors still contribute a certain noise level.
    Hence, the distance limit considered in \cite{hughes_qkd} still holds true.
    
    Our wavelength-multiplexed-based communication scheme shows advantageous behavior as compared to other strategies of increasing key rates.
    One possible scheme of increasing the secure key rate would be to improve the heralding efficiency, i.e., the ratio of coincident counts to singles counts, at the source \cite{anwar2020entangled}.
    This could be achieved by investing a tremendous experimental and technological effort improving the detector efficiencies.
    However, any potential benefit from increasing the heralding efficiency at the source is drowned by the dominant factor of link losses in any realistic QKD implementation.
    Another conceivable way of improving the key rate would be to use intensity-multiplexing devices \cite{Matthews2016} (i.e. $n$-port beam splitters).
    This corresponds to probabilistically distributing the photons on to multiple polarization analyzation stages.
    For high photon count rates (low losses), such a scheme could be used to partially increase the maximal achievable count rate governed by the dead time of the detector.
    However, in most realistic scenarios, losses lead to rather low count rates and limitations do to the detectors' dead times play no significant role \cite{neumann}.
    Moreover, contrary to our deterministic separation into wavelength channels, such a scheme does not allow to overcome the timing resolution limit set by the detectors. 
    The reason is that by increasing the photon-pair rate the probability of detecting multiple clicks on both receiver sides increases and Alice and Bob have no means to identify the actually correlated pairs, they are facing the same problem as without the probabilistic splitting being in place.
    Therefore, probabilistic multiplexing does not solve the problem of saturation in time.

\begin{acknowledgments}
    The authors thank Sebastian Ecker and Thomas Scheidl for useful comments and discussions.\\
    We acknowledge European Union's Horizon 2020 programme grant agreement No. 857156 (OpenQKD) and the Austrian Academy of Sciences. 
    We also gratefully acknowledge financial support from the Austrian Research Promotion Agency (FFG) Agentur für Luft- und Raumfahrt (FFG-ALR contract 844360 and 854022).
\end{acknowledgments}

    \textbf{Author contributions:}\\
    R.U. conceived the idea for the experiment.
    J.P. built the setup and conducted the measurement with the help of L.A.
    L.B. contributed to the data processing and analysis.
    J.P. and M.B. prepared the first draft of the manuscript and all authors contributed to writing the paper.
    R.U. and M.B. supervised the work.

\bibliography{apssamp.bib}
\bibliographystyle{apsrev4-1}

\end{document}